\documentclass[11pt,twoside,a4paper]{article}

\usepackage{epsfig}
\usepackage{graphics}

\usepackage{amsmath,amsthm,amsfonts,amssymb}

\renewcommand{\section}[1]
{
\vspace{10pt}
\rule{460pt}{.5pt}
\addtocounter{section}{1}
\setcounter{subsection}{0}
\setcounter{subsubsection}{0}
\noindent
{\large\bf \thesection {\ \ #1} \\}
\vspace*{-20pt} \\
\rule{460pt}{.5pt}
\vspace{-10pt}
}

\renewcommand{\subsection}[1]
{
\vspace{7pt}
\addtocounter{subsection}{1}
\setcounter{subsubsection}{0}
\noindent
{\large\bf
\thesubsection
{\ #1}}
\vspace{3pt}
}

\renewcommand{\subsubsection}[1]
{
\vspace{7pt}
\addtocounter{subsubsection}{1}
\noindent
{\large\it
\thesubsubsection {\ #1}}
\vspace{3pt}
}

\renewenvironment{thebibliography}[1]
    {\begin{list}{[\arabic{enumi}]}
    {\usecounter{enumi}
     \setlength{\parsep}{0pt}
     \setlength{\itemsep}{0pt}
     \settowidth{\labelwidth}{#1.}
     }}{\end{list}}

\begin{document}

\thispagestyle{empty}
\markboth{{\scriptsize \rm G.~Mavromanolakis}}
{{\scriptsize \bf Neural networks technique based signal-from-background separation ...}}
\pagestyle{myheadings}


\fontsize{11pt}{13.2pt}
\usefont{T1}{cmr}{m}{n}
\selectfont

\def\thefootnote{\fnsymbol{footnote}}


\vspace*{-45pt}
UA-NPPS/03/2003 \hfill {\sf \large hep-ex/0303021}


\vspace*{10pt}

{\Large \bf Neural networks technique based signal-from-background separation
and design optimization for a W/quartz fiber calorimeter
}

\vspace{15pt}
G.~Mavromanolakis~\footnote[1]
{
email: {\tt gmavroma@mail.cern.ch}
}

{\em \small University of Athens, Physics Department\\
Nuclear and Particle Physics Division\\
Panepistimioupoli Ilisia, 15771, Athens, Greece}

\vspace{15pt}
{\small
We present a signal-from-background separation study based on neural networks technique applied to
a W/quartz fiber calorimeter. Performance results in terms of signal efficiency and
improvement of the signal-to-background ratio are presented. We conclude that by using
neural networks we can efficiently separate signal from background and achieve a
signal enhancement over the background of the order of several thousands at high efficiency.
}

\section{Introduction}

Neural networks are widely used in scientific and commercial applications due to their generally better
performance compared to traditional statistical approaches and their relatively simple operation principle.
In high energy physics domain they
are commonly used in various pattern recognition problems
e.g. for quark and gluon jet identification~\cite{ref:lonnblad90,ref:lonnblad91,ref:graham95} or
in top quark~\cite{ref:holmstrom95} and Higgs boson searches~\cite{ref:maggipinto97,ref:hultqvist95},
track finding~\cite{ref:denby88,ref:peterson89,ref:gyulassy91},
triggering~\cite{ref:lindsey92}-\cite{ref:busson98},
data mining, and general classification tasks.
(Introduction to neural networks and review of their applications can be found
in~\cite{ref:csc91}-\cite{ref:horn97}.)

In this paper we present a signal-from-background separation study based on neural networks technique applied
to a W/quartz fiber calorimeter. The performance in terms of
signal efficiency and improvement of the signal-to-background ratio, and for various
calorimeter depths, read-out frequencies and total number of channels is presented.

The paper is organized as follows: in section~2  we give a general introduction
 on neural networks and related topics. The detector and its physics objectives are described in section~3.
Section~4 contains the analysis steps and the results. We summarize and conclude in section~5.

\section{Neural networks}

\subsection{General}

A neural network (NN) is a simplified mathematical structure inspired from the real biological neural networks
and their way of learning from experience, acquiring knowledge and solving problems.
Their basic units are the neurons, which are interconnected through synapses and exchange signals.
In general, a neuron produces an output signal which is depending on the signals it receives from the other
neurons. Of great importance is the fact that the output signal is non-linearly dependent to the input, whereas
the input signal is approximately the linear sum (the ``coefficients'' are determined by the synaptic strengths)
of all the  signals that are received by the neuron simultaneously.
The human brain consists of $O(10^{12})$ neurons, where each neuron is connected to a number, from
$O(1)$ to $O(10^{5})$, of other neurons. The whole structure is of immense complexity and plasticity and thus
 ability.

An NN has the basic concepts of a real biological neural network (neuron, connection strength,
input linearity, output non-linearity) but in a much more conservative level of complexity.
The neuron is a mathematical entity which has a real value depending on the connection strengths
(weights) and the values of the other neurons with which it is connected. The non-linear function
that relates the output from the neuron with the weights and the inputs to the neuron is usually called
{\em activation function} and has a simple sigmoid shape bound to values in the interval [0,1] or [-1,1].

A neural network can be {\em structured} or {\em self-organized}.
In a common structured topology the neurons are organized in layers. The {\em input layer}, from where
the NN is fed with the input variables of the problem to be solved, followed by a number of layers, the so-called
{\em hidden layers}, and finally the {\em output layer}.
 In the case where the information flow is in one direction only, from the input layer towards
the output layer, we speak about {\em feed-forward} neural networks. For bidirectional connections
we have {\em feed-back} NNs. When the output layer is fed back into the input layer then we have
{\em recurrent networks}.

An NN is trained by a set of examples which are representative of the problem. During training the weights, which
are the state variables of the network and determine its behavior, are adapted to the presented examples.
In such a way the NN acquires knowledge of the rule which produced the examples and then it can generalize. Generalization
means that the network can be used on real events and perform the task that was trained for. Its generalization
ability is tested usually with a set of test events that we know their properties, with this procedure
we validate the network's performance and afterwards it is ready to be used on real events.

The training procedure can be {\em supervised} or {\em unsupervised}.  Supervised training is accomplished
by presenting input-output pairs to the NN. The NN calculates its output, according to its weights,
which is compared to the given desirable output. The weights are updated in such a way that the
NN produces output as close as possible to the desirable one. In unsupervised training
the network receives only a set of input examples without any output labeling. In this case the task is
to detect some, yet unknown, structure that the real events may have.

In the following we concentrate on one of the most commonly used feed-forward neural networks,
the {\em multilayer perceptron}. It is usually being trained with the {\em back-propagation algorithm} to perform
event classification tasks. We describe its architecture and present the mathematical background of the training
algorithm. First we discuss in brief the problem of {\em pattern recognition}.

\subsection{Pattern recognition}

Pattern recognition for event classification is a common problem encountered in high energy physics
(e.g. particle identification by its shower dimension and shape, electron/hadron discrimination,
trigger signal generation, quark and gluon jet classification etc.).
In all cases, the problem consists in defining a procedure that should be followed in online or offline analysis,
and will be able to recognize events and categorize them based on their features.
A feature or pattern, is the set of properties that a class of events is characterized of,
and with which this class can be possibly discriminated by the other ones.
The difficulty is to reveal characteristic patterns from the measured quantities.
Therefore the whole procedure is to find and properly use quantities or combination of them, that will
allow correct event classification.
In an approach without NNs, one usually tries to perform classification by imposing a set of one-dimensional
cuts on various selected measured variables, that may characterize the events of interest.
Such cuts are usually
determined by examining single variable probability distributions for the events of interest and  for the others
(see~e.g.~\cite{ref:fukunaga90}).
When the complexity of the problem is increasing, better results can be obtained by employing methods that
could simultaneously exploit correlations among variables. In this case a procedure with high degree
of parallelism is required. The neural networks are built by concept with this kind of parallelism (neurons), and therefore
they are usually used on pattern recognition tasks with outstanding performance.

\subsection{Multilayer perceptron}

\begin{figure}[tb]
\centering
\epsfxsize=250pt
\epsffile{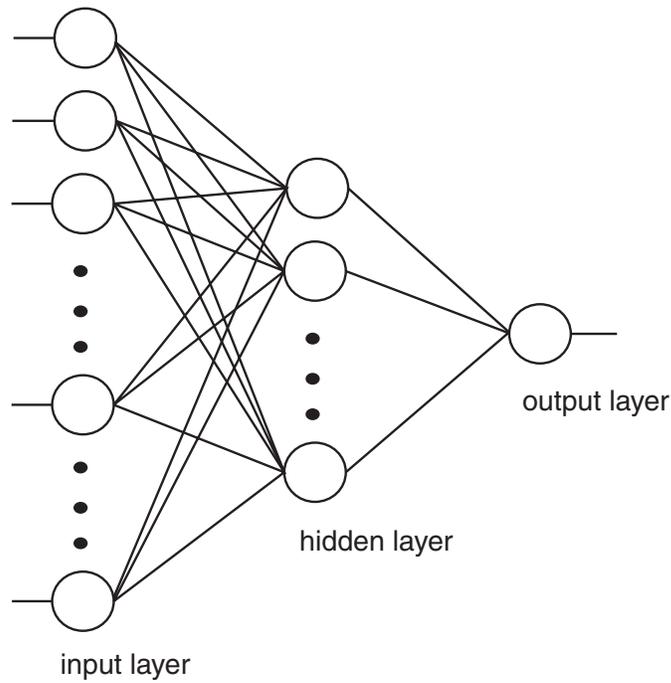}
\caption{a multilayer perceptron with one hidden layer.}
\label{fig:mlp}
\end{figure}

One of the most widely used NNs for pattern recognition tasks is the multilayer perceptron.
It is a feed-forward network with its neurons arranged in input, hidden and output layers~(fig.~\ref{fig:mlp}).
Every neuron is connected to all neurons of the preceding layer and of the next one, whereas
there is no connection between neurons belonging to the same layer.
Every neuron receives inputs from all
neurons of the preceding layer, and sends its output to all ones in the next layer.
All neurons obey to this connectivity scheme. Every connection is associated to a weight parameter that will
be optimized during the training phase.
The neuron's output is given by the activation function $f(x)$, a sigmoid function like eq.~\ref{eq:tanh},
that has values
bound to the interval $[-1,1]$, or like eq.~\ref{eq:exp}, with values
bound to the interval $[0,1]$. The parameter $T$, usually called temperature of the network, determines
how steep the sigmoid function is, the lower $T$ the steeper sigmoid. In general the final performance
does not depend on $T$, a value equals to 1 is the usual choice.
The input layer consists of a number of neurons which is equal
to the $N_{input}$ input variables that will be used for the event classification.

Presenting these in mathematical form,
we define $I_{i}^{k}$ to be the input to the $i$th neuron in the $k$th layer, which consists of
$N^{k}$~(\footnote[1]{in this notation $k$ serves as an index and not as an exponent value}) neurons,
$O_{i}^{k}$ to be the output of the $i$th neuron in the $k$th layer and $w_{ji}^{k-1}$ to be the weight
of the connection between the $j$th neuron in the $(k-1)$th layer and the $i$th neuron in the $k$th layer.
Then :
\begin{eqnarray}
\label{eq:input}
  I_{i}^{k} = \sum_{j=1}^{N^{k-1}} w_{ji}^{k-1} O_{j}^{k-1} & , & k>1
\end{eqnarray}
\begin{eqnarray}
\label{eq:output}
O_{i}^{k} = f(I_{i}^{k} + w_{i0}^{k}) & , & k>1
\end{eqnarray}
where $f(x) \equiv {\rm sigmoid \: function}$,
e.g.
\begin{eqnarray}
\label{eq:tanh}
f(x) = \tanh(x/T)
\end{eqnarray}
or
\begin{eqnarray}
\label{eq:exp}
f(x) = \frac{1}{1 + e^{-x/T}} &,
\end{eqnarray}
and $w_{i0}^{k}$ is a so-called {\em bias} or {\em threshold} of the neuron. During the training phase it is
simply considered as a weight and it is optimized in a same way.

For $k=1$ (input layer) it is :
\begin{eqnarray}
O_{i}^{1}  \equiv I_{i}^{1} = {\rm input \: variables} & , & i=1, ..., N_{input}
\end{eqnarray}

The output layer may consist of one neuron and in this case the network is suitable for two classes separation
problem. In general there may be $N_{output}$ neurons when the events should be categorized in
$2^{N_{output}}$ classes. In this case the $k$th bit in a binary representation of the class number is determined
by the value of the $k$th output neuron. Alternatively, the individual $N_{output}$ neurons could be associated
with particular classes among of $N_{output}$ possible choices.

Concerning the number of hidden layers, a general rule is that no more than two hidden layers should be
needed~\cite{ref:csc91}.
Experience shows that whether a multilayer perceptron with two hidden layers performs
satisfactorily or not, then by adding to it more hidden layers its performance is not improved.
In most cases only one hidden layer is sufficient. In fact for problems of function
approximation, it has been shown~\cite{ref:hornik89,ref:cybenko89,ref:hornik90}
that a linear combination of sigmoids can approximate any
continuous function of one variable or more. Obviously (eq.~\ref{eq:input},~\ref{eq:output}) this can be
achieved with a network with one hidden layer.

Concerning  the number of hidden neurons that a NN should have, there is no general rule that could
suggest which number is optimal. In general the number of hidden neurons should be large enough
to ensure a high degree of classification, and small enough to ensure a high degree of generalization.
If there are too many hidden neurons, the network tends to learn to recognize only the set of examples
that were used for its training. Therefore its generalization ability is poor and it does not perform
well on a test sample or/and on real events. If the number of neurons is too small, the network performs
bad since during the training phase it is unable to learn to classify.

In the following we describe the back-propagation algorithm which is usually used for training
a multilayer perceptron.

\subsection{Back-propagation algorithm}

The back-propagation algorithm is a supervised training method often used with various modifications
for training feed-forward networks. A set of examples is presented to the network which determines
its output according to its state variables, the weights. The output is compared to
the desired target output that each example is labeled with. By the difference between the network's output
and the target one, a so-called {\em error} or {\em cost function} is determined. It is actually a function
of the weights, and the algorithm should update and modify these free parameters in such way that
the error function is minimized, or in other words the network's output is as close as possible to the
target output. The minimization is done with a gradient descent method with respect to the weights.

We describe the algorithm following the notation given in the previous subsection.
We consider a multilayer perceptron that is composed of $H+2$ layers
(an input layer, $H$ hidden ones and an output layer). There are $N_{weights}$ parameters
(connection weights and thresholds are simply called weights)
that should be optimized, where
\begin{eqnarray}
\label{eq:Nweights}
N_{weights} = \sum_{k=2}^{H+2}( N^{k} + N^{k}N^{k-1})
\end{eqnarray}

The error function is defined as
\begin{eqnarray}
\label{eq:error_function}
E(t) = \frac{1}{2} \sum_{n=1}^{N_{examples}} (O(n;t) - T(n))^2
\end{eqnarray}
where $O(n;t)$ is the output of the network corresponding to
the $n$th example after the minimization process has been iterated $t$ times, and $T(n)$ is the target output
associated with the example. For simplicity we consider the case of event classification between
two classes (signal associated with target output $T=1$ and background with $T=0$),
and so the output layer has only one neuron (denoted as $O^{H+2}$ or simply $O$).

The weights are modified after each iteration according to
\begin{eqnarray}
w_{ji}^{k}(t+1) = w_{ji}^{k}(t) + \Delta w_{ji}^{k}(t+1) & , & k=1, ..., H+1
\end{eqnarray}
with
\begin{eqnarray}
\Delta w_{ji}^{k}(t+1) = -\eta\frac{\partial E(t)}{\partial w_{ji}^{k}(t)} + \alpha \Delta w_{ji}^{k}(t)
& , & k=1, ..., H+1
\end{eqnarray}
where the parameter $\eta$ ({\em learning factor}) determines the step of change, and thus the training rate,
and $\alpha$ ({\em momentum coefficient}) is a smoothing parameter that helps the method
to avoid getting stuck around the local minima of the error function.

The partial derivative of the error function is given by the recurrent relations (event and iteration
indices are not shown for convenience)
\begin{eqnarray}
\label{eq:recurrent1}
\frac{\partial E}{\partial w_{ji}^{k}}
=
D_{i}^{k+1}
\cdot
\left.\frac{\partial f(x) }{\partial x}\right|_{x = I_{i}^{k+1} + w_{i0}^{k+1}}
\cdot
O_{j}^{k} & , & k=1, ..., H+1
\end{eqnarray}
with
\begin{eqnarray}
\label{eq:recurrent2}
D_{i}^{k+1}
=
\sum_{l=1}^{N^{k+2}}
D_{l}^{k+2}
\cdot
\left.\frac{\partial f(x) }{\partial x}\right|_{x = I_{l}^{k+2} + w_{l0}^{k+2}}
\cdot
w_{il}^{k+1} & , & k=1, ..., H
\end{eqnarray}
and
\begin{eqnarray}
D_{}^{H+2}
=
O_{}^{H+2} - T
\end{eqnarray}

As can be seen in equations~\ref{eq:recurrent1} and~\ref{eq:recurrent2}, the update information flows back
from the last layer to the previous one (back-propagated).
The weights of the neural network can be updated either after each example presentation ({\em online training}),
or after the whole set of examples has been processed ({\em batch training}).

In general, we anticipate to reach a state in which the update differences
$\Delta w_{ji}^{k}$ are zero or close enough to it.
After the training phase, we validate the quality of classification task
that the network achieved with a test sample of events (generalization ability).
The neural network is now ready to classify real events, that has never seen before, with known classification efficiency.

The main disadvantage of the algorithm presented above, often called as the {\em standard back-propagation method},
is that it converges very slowly to the minimum of the error function. Besides this convergence time
grows very fast with the complexity of the problem and the size of the network.
There are a couple of decades of different methods, that are based on the core features of the standard one,
which achieve to speed up the overall performance significantly, at least in most cases, by e.g. defining
a different error function, changing the learning factor or the network's temperature with iteration, using second order
derivatives or adding extra terms, or even by allowing the network to generate or destroy neurons (self-generation)
in order to reach better performance. For a rigorous description and references on several
methods consult~\cite{ref:csc91}.

For any training method, concerning the number of example events ($N_{examples}$) that one should use on training,
a general rule is that it should be one or two orders of magnitude larger than the number of weights
($N_{weights}$) which are composing the neural network. This is imposed by the fact that the
generalization ability depends mainly on the ratio $N_{weights}/N_{examples}$. Actually, for a multilayer perceptron
with one hidden layer it has been shown that the generalization error is of
$O(N_{weights}/N_{examples})$~\cite{ref:generalization_error}.

Of special care is the question on when the training phase should be stopped. The answer to this
is, whenever there is indication of network {\em over-training}.
This is the case where the network
starts learning to recognize only the set of training examples,
and as a consequence, losing its generalization ability. This can be avoided by performing training and testing in parallel
and inspect the evolution of the error function for both samples with respect to the iteration of the training algorithm.
When the value of the error function on the testing examples starts to increase, while on the training ones it continues
decreasing, there is danger of over-training if we continue the training procedure.
There is no rule that associates maximum number of iterations with over-training or with best training.
Generally, this depends on the complexity of the classification problem, the complexity of the neural network,
the training method and its learning factor. If a network does not show satisfactory performance before
over-training occurs, then it means that the approach to that classification problem with neural network technique
unfortunately fails.

We close this introduction to neural networks by adding another, somehow abstract but at the same time hopefully
more clear,
description of the neural networks approach to pattern recognition problems. The whole procedure can be considered
as a fitting problem of  $N_{examples}$ ``points'' on an $N_{input}\times N_{output}$ -dimensional ``plane'' with
a ``curve'' (neural network) of $N_{weights}$ free parameters. The best fitting ``curve'' (trained neural network),
is simply represented by a function parameterized by the weights (free parameters that were optimized)
that receives $N_{input}$ arguments (the network's input variables) and
returns a value (results in the output layer) which approximates the corresponding ``point's position''
(classification).

\section{The CASTOR calorimeter}

\subsection{Motivation} 

\begin{figure}[tbp]
\centering
\epsfxsize=460pt
\epsffile{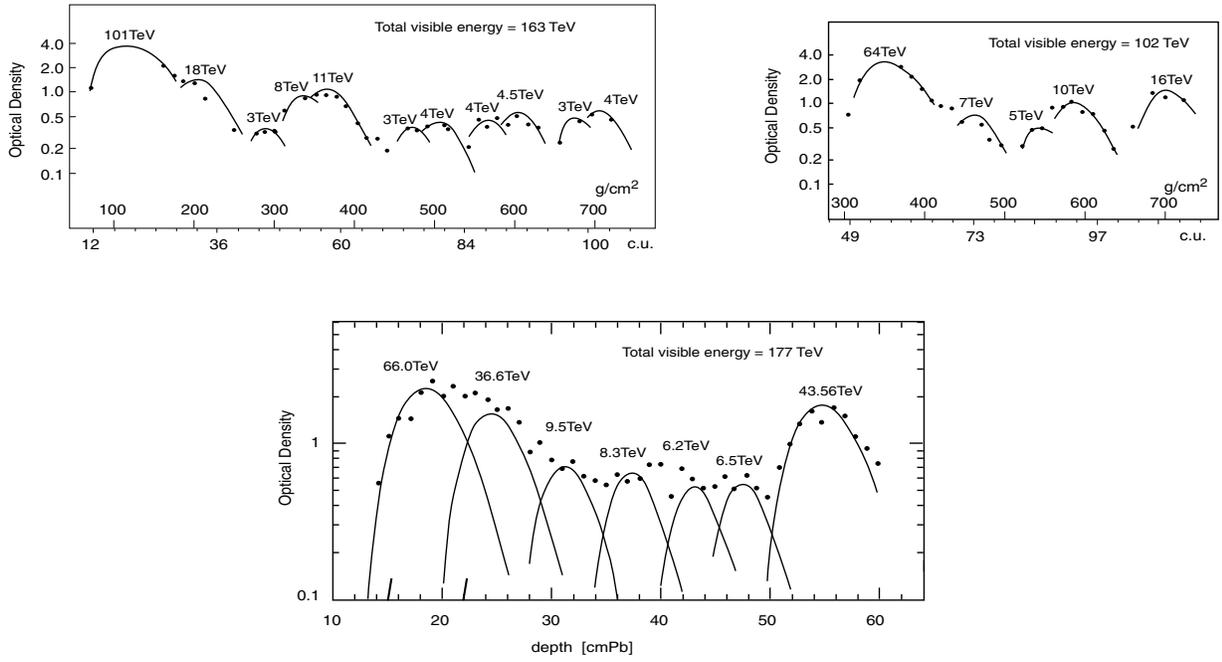}
\caption{longitudinal profile of ``long penetrating objects'' observed in multilayer lead-emulsion chambers
(top two plots from~\cite{ref:buja81}, bottom one from~\cite{ref:arisawa94}).}
\label{fig:longpenetrating}
\end{figure}

Cosmic rays experiments have detected events with unusual properties, the so-called {\em Centauro}
events \cite{ref:lattes80}-\cite{ref:arisawa94},
which exhibit small particle multiplicity compared to normal hadronic events,
complete absence or strong suppression of
the electromagnetic component ($N_{hadrons}/N_{\gamma} \gg 1$, $E_{hadrons}/E_{\gamma} \gg 1$)
and very high $<p_t>$. Furthermore, a number of
hadron-rich events are accompanied by a strongly penetrating component observed
in the form of halo~\cite{ref:baradzei92}, strongly penetrating clusters~\cite{ref:hasegawa96} or
long-living cascades, whose transition curves exhibit a characteristic form
with many maxima and slow attenuation (e.g.~fig.~\ref{fig:longpenetrating})~\cite{ref:arisawa94,ref:buja81}.
These events can not be explained in terms of statistical deviation from conventional hadronic
physics~\cite{ref:tamada97,ref:barroso99,ref:tamada99}.

According to a phenomenological
model~\cite{ref:asprouli94,ref:panagiotou92,ref:panagiotou89},
the Centauros are considered to be the
products of hadronization of a deconfined quark-matter fireball formed in
nucleus-nucleus collisions in the upper atmosphere. The ``long penetrating objects''
usually accompanying them are assumed to be long-lived strangelets, that may have been formed
because of a mechanism of strangeness separation ~\cite{ref:greiner88,ref:greiner87} of
the fireball's strange quark content and are emitted during fireball's hadronization.
In a similar way, this kind of events and particles may be produced in Pb+Pb collisions at the
LHC ($\sqrt{S} = 5.5$~TeV/nucleon) from the hadronization of a Quark Gluon Plasma state formed in the beam
fragmentation region.

\subsection{Detector description}

\begin{figure}[tb]
\centering
\epsfxsize=350pt
\epsffile{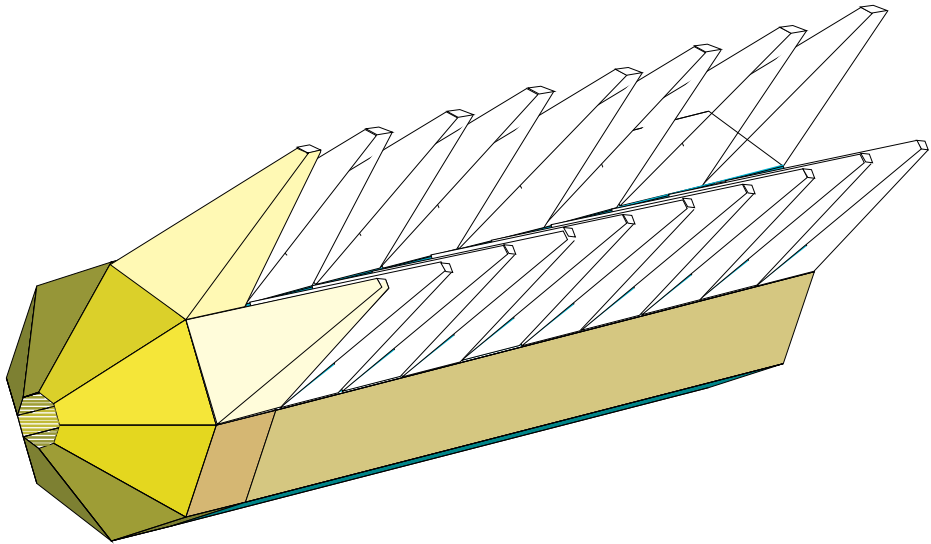}
\caption{schematic view of the CASTOR calorimeter, some of its air-core lightguides are shown.}
\label{fig:calo_view}
\end{figure}

The CASTOR detector~\cite{ref:castor99,ref:castor97} is a Cherenkov effect based sampling calorimeter with
a tungsten absorber and quartz fibers as active material. The signal is the Cherenkov light produced by
the shower charged particles traversing the fibers. The calorimeter is azimuthally divided
in 8 octants and longitudinally segmented in layers. Each absorber layer is
followed by a number of quartz-fiber planes, altogether consisting a W-fiber layer.
The W-fiber layers have 45$^{\circ}$
inclination with respect to the beam axis to achieve maximum light production.
The calorimeter consists of several channels per octant. One channel consists of a number of consecutive
W-fiber layers, the signal of which is collected and transmitted to its
corresponding photomultiplier through an air-core lightguide (fig.~\ref{fig:calo_view}).
The calorimeter is proposed as a very forward detector of the ALICE~\cite{ref:alice_proposal}
or CMS~\cite{ref:cms_proposal} experiments at the LHC
covering the pseudorapidity range  $5.46 \leq\eta\leq 7.14$. Its main
objective is to search in the baryon rich, very forward rapidity region of central
Pb+Pb collisions  for unusual events and ``long penetrating objects'', assumed to be strangelets,
by measuring the hadronic
and electromagnetic energies and the hadronic shower's longitudinal profile on an event-by-event mode.

Detailed simulation studies of the performance of the CASTOR calorimeter have
been done~\cite{ref:gm1,ref:gm2}. The calorimeter shows linear response to
electrons and hadrons, satisfactory energy resolution and very narrow visible
transverse size of electromagnetic and hadronic showers, a property that derives
from the detector's operation principle, based on the Cherenkov effect, which makes
such calorimeters sensitive essentially only to the shower core~\cite{ref:britz95,ref:gorodetzky95}.
Concerning the lightguides, their shape, dimensions and inner walls have been studied
\cite{ref:gm3} and are optimized for better light transmission efficiency.

In this point we wish to refer the ALICE-PMD (Photon Multiplicity Detector)~\cite{ref:alice_pmd},
which covers the pseudorapidity region
$1.8 \leq\eta\leq 2.6$  and is dedicated to the measurement of photon and charged particle multiplicities.
It is designed for
Centauro events related research but from a different viewpoint.
Its objective is to detect possible large non-statistical fluctuations on an event-by-event basis
which is the primary signature of the formation
of Disoriented Chiral Condensate (e.g.~\cite{ref:aggarwal98}).
Also, concerning strangelets research in the central rapidity region covered by
the ALICE barrel detector, a similar search as has been performed on recent
fixed target heavy-ion experiments (NA52 \cite{ref:appelquist96,ref:ambrosini96} at CERN-SPS
and E864 \cite{ref:E864_97_1,ref:E864_97_2,ref:E864_99} at BNL-AGS), aiming on detection of particles
with low charge-to-mass ratio,
is forseen~\cite{ref:coffin97,ref:alice_proposal} using the central tracking system.
(For a rigorous review on strange quark matter searches consult~\cite{ref:klingenberg99}).

\subsection{``Long penetrating objects''}

The hypothesis that the long penetrating objects may be strangelets is supported by simulations~\cite{ref:gladysz97}
which show that the passage of a strangelet through matter produces shower which is slowly attenuated,
long penetrating and has a longitudinal profile with many maxima structure, as observed in cosmic rays
experiments~\cite{ref:arisawa94,ref:buja81}. The passage of strangelets through the CASTOR calorimeter
has been also simulated~\cite{ref:angelis99} and
the analysis of the results has shown that the signal can be easily distinguished from the hadronic
background for strangelets with energy greater than 20 TeV. Nevertheless, strangelets with a such high energy
are expected to be boosted in high rapidity (and thus pseudorapidity) and as a consequence it is very likely that
they pass outside the detector's coverage. In addition, for the identification of lower energy strangelets,
a calorimeter with very high read-out frequency is required, which is not feasible. In this study we present
a sophisticated method based on neural networks technique for the separation of the low energy strangelets signal
from the hadronic background.

\section{Signal-from-background separation analysis}

The forward region ($5.46 \leq\eta\leq 7.14$) covered by the CASTOR calorimeter receives $200 \pm 11$~TeV
per central Pb+Pb collision at $\sqrt{S} = 5.5$~TeV/nucleon, carried by about 2000 particles (event
generated by HIJING~\cite{ref:hijing}).
This amount of energy is associated with the conventional hadronic events, treated as background
for CASTOR's research interests. We make the basic assumption that in case that a strangelet is produced in a
Pb+Pb collision, the energy not carried by it is going into conventional particle production as
described by the event generator.
In other words, although the detector may receive the same amount of energy, in the first case the event
should be considered as not interesting (background), whereas in the second one, as signal. The only
discriminating feature between them is the fact that the strangelet's passage through the detector
gives a shower with many maxima and negligible attenuation.

We study the case of a 5~TeV strangelet, an amount of
energy which corresponds to 2.5\% of the total energy per event that is received by the calorimeter,
a fact that makes the separation task not trivial because of mainly two reasons. First, the
characteristic pattern of the longitudinal development of the shower is weak and can be easily masked
and suppressed by fluctuations of the showers of the other hadrons. Second, due to the fact that the calorimeter
is divided in a reasonable (low) number of channels, and thus the characteristic many-maxima signal is likely to be
distributed to consecutive channels and undersampled. The situation would be even worse with a detector
which was not azimuthally divided. The 8 octants of the CASTOR detector are operating as stand-alone calorimeters
since the visible shower transverse size is very narrow~\cite{ref:gm1,ref:gm2}.

We should also mention the fact that we must be able to cope with an expected signal-to-background ratio
(in the raw data recorded)
of the order of 1/10000.
As it is shown in the followings, by using neural networks we can surpass these difficulties
and achieve very efficient separation of signal from background.

\subsection{Analysis steps description}

\begin{figure}[tbp]
\centering
\epsfysize=180pt
\epsffile{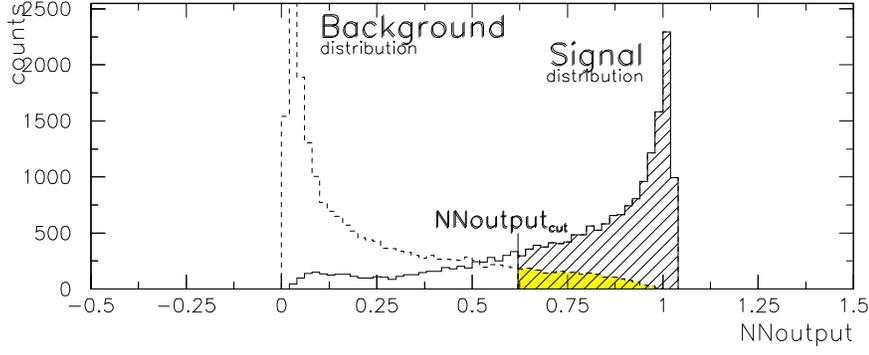}
\caption{typical signal and background distribution as a function of NNoutput. The hatched area contains
the $N_{signalNN}$ signal events that are above the selection cut NNoutput$_{cut}$. The colored area contains
the contaminating $N_{signallikeNN}$ background events.}
\label{fig:typicalNNoutput}
\end{figure}

Since the calorimeter is composed of 8 octants which are operating and read-out as individual
detectors, the task of event classification consists in disentangling an octant which contains
the signals of a strangelet and the accompanying particles (signal event) from an octant which contains
only conventional signal (background event). The neural network is fed with input variables that are
the responses of the channels of the octant and should provide a discriminating value (NNoutput) which
is close to 1 for signal events and 0 for background ones.
A typical distribution of signal and background events as a function of NNoutput is
shown in fig.~\ref{fig:typicalNNoutput}. By applying a suitable cut (NNoutput$_{cut}$) we can select
a subset which contains sufficiently high number of signal events and low number of contaminating background.
We define the followings:

{\em signal efficiency}, $\epsilon_{s}$, which represents the probability of classification of a real signal event as signal,
\begin{eqnarray}
\label{eq:signal_eff}
\textit{signal efficiency: } \epsilon_{s}=\frac{N_{signalNN}}{N_{signal}}
\end{eqnarray}
where $N_{signalNN}$ is the number of events that have been selected out of $N_{signal}$ signal events
due to the fact that they produce a NNoutput which satisfies the condition to be greater than the imposed cut value
NNoutput$_{cut}$,

{\em contamination}, $\epsilon_{b}$, which represents the probability of misclassification of a background event as signal,
\begin{eqnarray}
\label{eq:contamination}
\textit{contamination: } \epsilon_{b}=\frac{N_{signallikeNN}}{N_{background}}
\end{eqnarray}
where $N_{signallikeNN}$ is the number of events that, although they belong to the set of the $N_{background}$
background events, they produce a NNoutput value which is greater than NNoutput$_{cut}$ and thus they are
misclassified as signal events,

{\em signal enhancement}, $\frac{\epsilon_{s}}{\epsilon_{b}}$, which represents the factor of improvement
of the signal-to-background ratio,
\begin{eqnarray}
\label{eq:signal_enhanc}
\textit{signal enhancement} = \frac{\epsilon_{s}}{\epsilon_{b}}
\end{eqnarray}
The parameters $\epsilon_{s}$, $\epsilon_{b}$, $\frac{\epsilon_{s}}{\epsilon_{b}}$  are determined at the training and
testing phase of the network and quantify its performance. Then by applying the trained network on e.g. a set of
raw data that was taken during experiment run and is assumed to have a signal-to-background ratio
%
$S/B=\frac{N_{signal}}{N_{background}}$,
we can result in a subset of data to be further analyzed with a signal-to-background ratio $(S/B)_{NN}$
enhanced by a factor of $\frac{\epsilon_{s}}{\epsilon_{b}}$,
\begin{eqnarray}
\label{eq:S/B_NN}
(S/B)_{NN}=\frac{N_{signalNN}}{N_{signallikeNN}}
=\frac{\epsilon_{s}}{\epsilon_{b}}\cdot\frac{N_{signal}}{N_{background}}=\frac{\epsilon_{s}}{\epsilon_{b}}\cdot S/B
\end{eqnarray}

For the construction of the neural networks, their training and testing we used the environment
provided by the MLPfit package~\cite{ref:mlpfit}, a tool with great functionality in development of multilayer perceptrons.
Other packages usually used are Jetnet~\cite{ref:jetnet}, SNNS~\cite{ref:snns}, NNO~\cite{ref:nno}.
A set of 10000 signal events and 10000 background ones,
which represents the calorimeter's octant simulated response to the interesting and non-interesting events as described above,
is used for training the networks. A second independent set composed by 10000+10000 events is used in the testing
phase. Several network architectures have been used with one hidden layer since by adding a second
layer the performance is not improved. Several training algorithms have been initially used, and as expected
without causing any significant change in the final network performance. We chose to work with the BFGS algorithm
(Broyden-Fletcher-Godfarb-Shanno) since it is fast, very efficient and reliable~\cite{ref:mlpfit,ref:bfgs}.

The general specifications of the calorimeter we used in this study are tabulated in table~\ref{tab:specifications}.
In order to investigate how the total calorimeter depth and the total number of channels influence
the signal detection efficiency, we studied 9 different configurations tabulated in table~\ref{tab:configurations}.
They can be categorized in three cases, according to total calorimeter depth or according to depth per
channel. In the following subsections we use the second categorization scheme.
The results are presented in terms of signal enhancement ($\frac{\epsilon_{s}}{\epsilon_{b}}$)
and signal classification efficiency ($\epsilon_{s}$).

\begin{table}[tb]
\centering
  \caption{calorimeter specifications.}
  \label{tab:specifications}
\vspace{10pt}
  {\small
  \begin{tabular}{ll}
    \hline\hline
                & \\
{\bf absorber:} & W ($\lambda_{I}$=10.0 cm, $X_0$=0.365~cm, density=18.5 gr/cm$^3$)\\
                & maximum 170 layers, 0.5 cm thick each (= 0.071 $\lambda_{I}$ after 45$^\circ$ inclination)                                   \\
                & \\
{\bf fiber:}    & quartz core (diameter 600 $\mu$m), hard plastic cladding (diameter 630 $\mu$m)\\
                & numerical aperture = 0.37 \\
                & 3 fiber planes per absorber layer ($\equiv$ 1 W-fiber layer)\\
                & \\
{\bf filling ratio:}&$\frac{fiber \ volume}{absorber \ volume}$ : 26.5\%\\
                & \\
{\bf channels:} & configurations with 7, 10, 15 consecutive W-fiber layers per channel \\
                & \\
\hline\hline
  \end{tabular}
  }
\end{table}

\begin{table}[tb]
\centering
  \caption{calorimeter configurations.}
  \label{tab:configurations}
\vspace{10pt}
  \begin{tabular}{c|rrr}
    \hline\hline
  $\lambda_{I}$'s(layers)     & \multicolumn{3}{c}{\# of channels(layers) per octant} \\
  per channel                 & \multicolumn{3}{c}{for calorimeter depth} \\
                              & $\sim 9.3 \lambda_{I}$'s & $\sim 10.5 \lambda_{I}$'s & $\sim 11.7 \lambda_{I}$'s \\
\hline
& & & \\
0.49 ( 7) & 19 (133) & 21 (147)  & 24 (168) \\
0.70 (10) & 13 (130) & 15 (150)  & 17 (170) \\
1.05 (15) &  9 (135) & 10 (150)  & 11 (165) \\
& & & \\
\hline\hline
  \end{tabular}
\end{table}

\subsection{Calorimeter of 1.05 $\lambda_{I}$ per channel}

\begin{figure}[tbp]
\centering
\epsfxsize=270pt
\epsffile{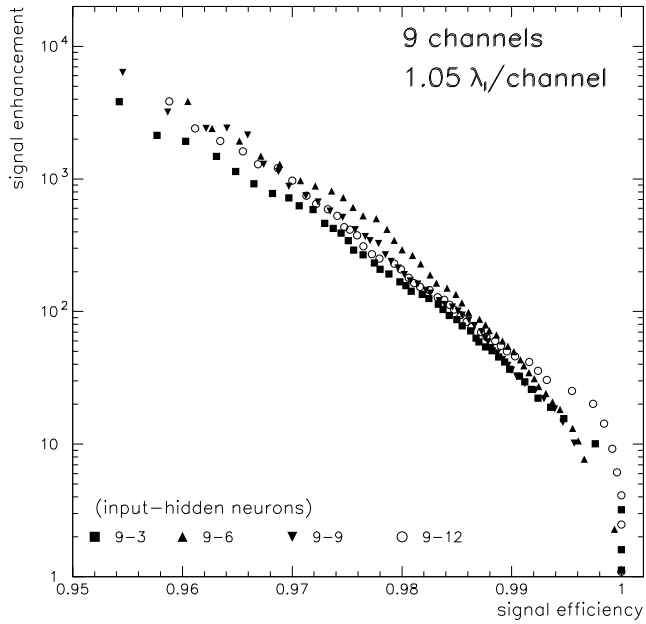}
\caption{signal enhancement ($\frac{\epsilon_{s}}{\epsilon_{b}}$) as a function of signal efficiency
($\epsilon_{s}$) for a calorimeter with 9~channels/octant and total depth of 9.45~$\lambda_{I}$'s.}
\label{fig:sumnnall9}
\end{figure}

\begin{figure}[tbp]
\centering
\epsfxsize=270pt
\epsffile{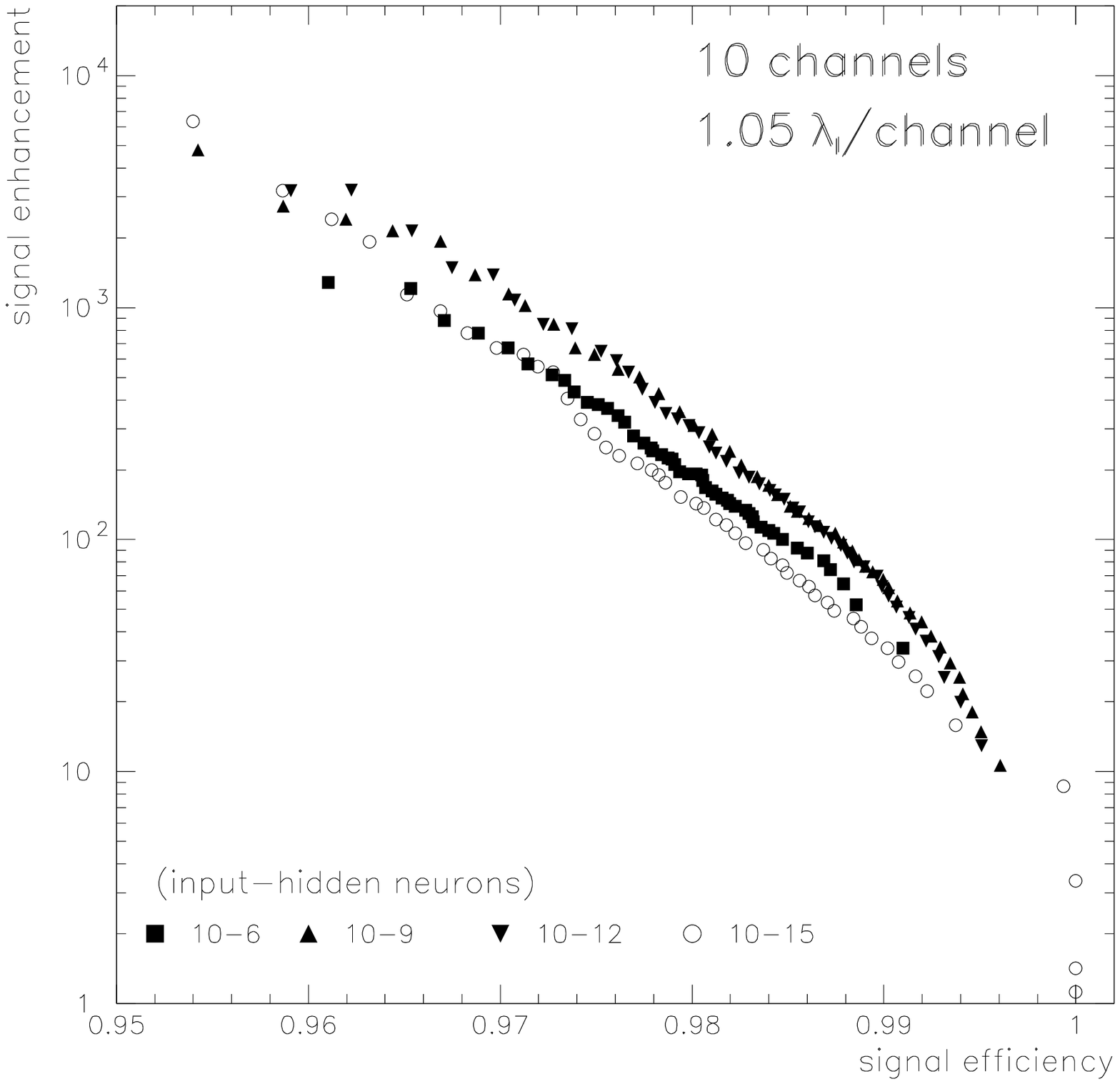}
\caption{signal enhancement ($\frac{\epsilon_{s}}{\epsilon_{b}}$) as a function of signal efficiency
($\epsilon_{s}$) for a calorimeter with 10~channels/octant and total depth of 10.50~$\lambda_{I}$'s.}
\label{fig:sumnnall10}
\end{figure}

\begin{figure}[tbp]
\centering
\epsfxsize=270pt
\epsffile{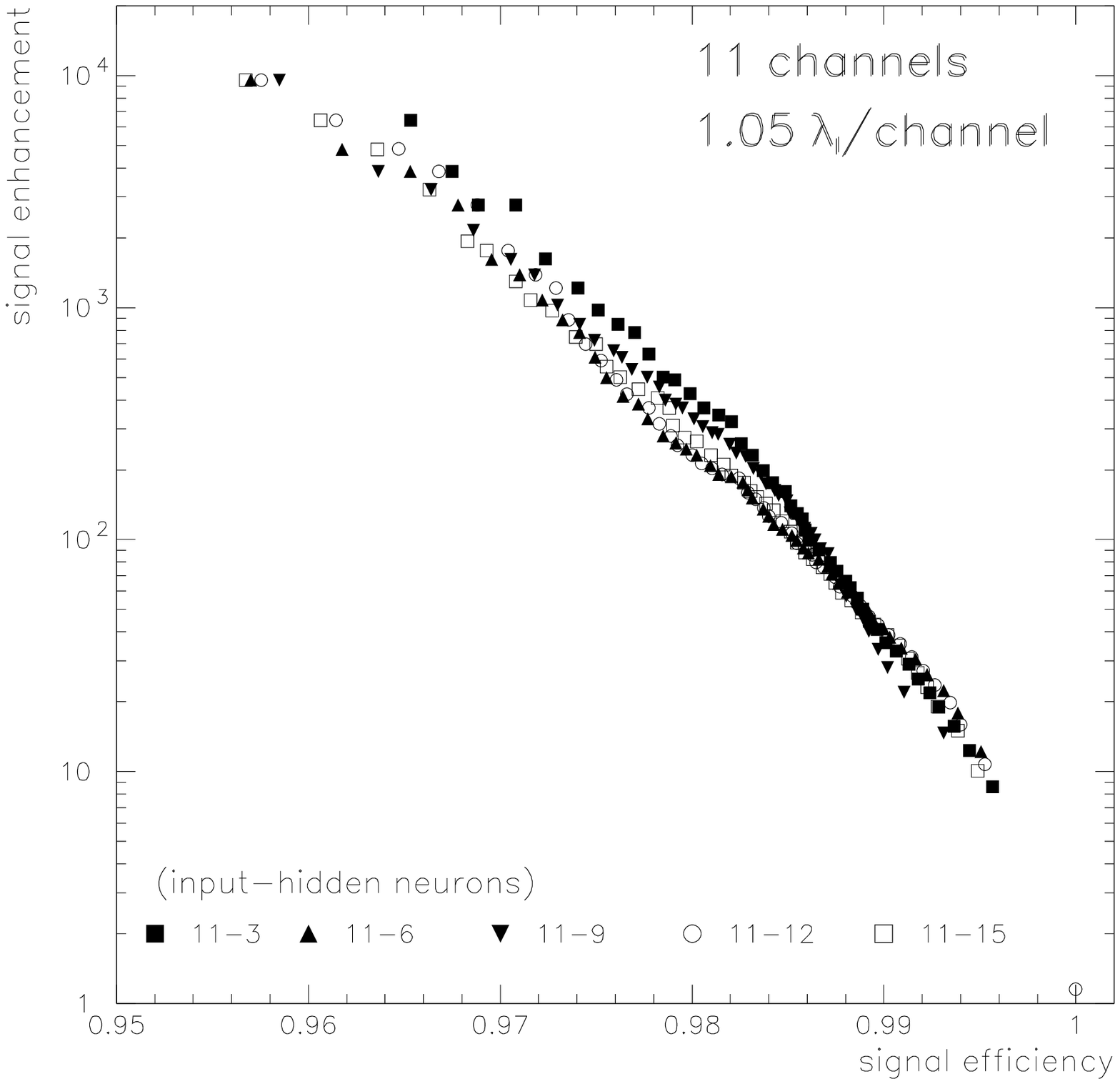}
\caption{signal enhancement ($\frac{\epsilon_{s}}{\epsilon_{b}}$) as a function of signal efficiency
($\epsilon_{s}$) for a calorimeter with 11~channels/octant and total depth of 11.55~$\lambda_{I}$'s.}
\label{fig:sumnnall11}
\end{figure}

We first consider the case where each channel consists of 15 consecutive W-fiber layers, which
correspond to 1.05 $\lambda_{I}$ per channel ($= 28.8~X_0$/channel). We studied calorimeters that are
composed of 9, 10 and 11 channels per octant (total depth is 9.45, 10.50, 11.55 $\lambda_{I}$'s, respectively).
Neural networks with various numbers of hidden neurons have been used. Their performances in terms
of signal enhancement ($\frac{\epsilon_{s}}{\epsilon_{b}}$) as a function of signal classification
efficiency ($\epsilon_{s}$) are depicted in figures~\ref{fig:sumnnall9},~\ref{fig:sumnnall10}~and~\ref{fig:sumnnall11}.
Although the performance varies among different configurations, for all cases a signal enhancement higher
than 1000 can be achieved at satisfactorily high efficiency.
The results from each neural network configuration are presented analytically in
table~\ref{tab:NNs_performance} at page~\pageref{tab:NNs_performance} and
discussed in subsection~4.5.


\subsection{Calorimeter of 0.70 $\lambda_{I}$ per channel}

\begin{figure}[tbp]
\centering
\epsfxsize=270pt
\epsffile{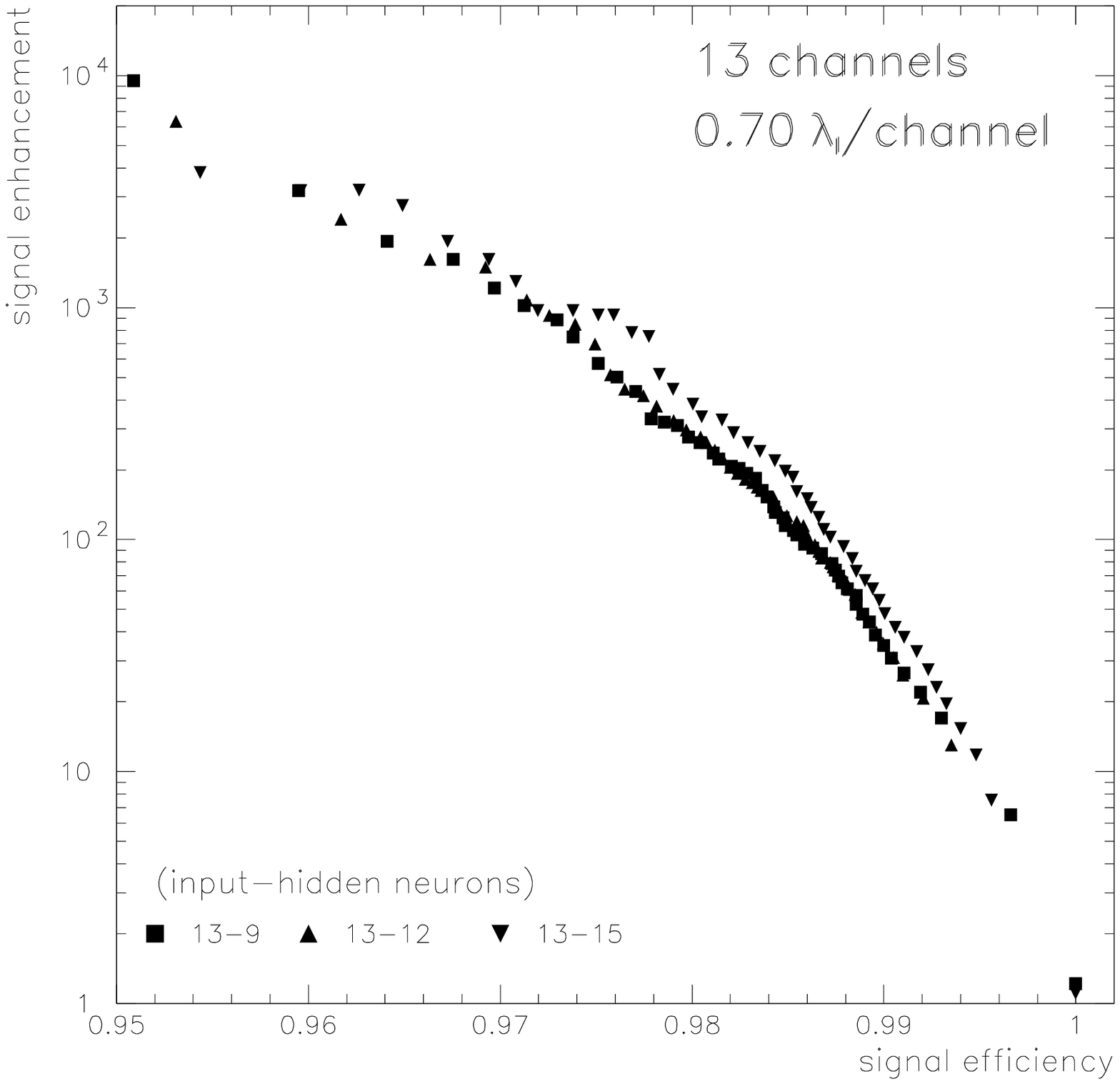}
\caption{signal enhancement ($\frac{\epsilon_{s}}{\epsilon_{b}}$) as a function of signal efficiency
($\epsilon_{s}$) for a calorimeter with 13~channels/octant and total depth of 9.10~$\lambda_{I}$'s.}
\label{fig:sumnnall13}
\end{figure}

\begin{figure}[tbp]
\centering
\epsfxsize=270pt
\epsffile{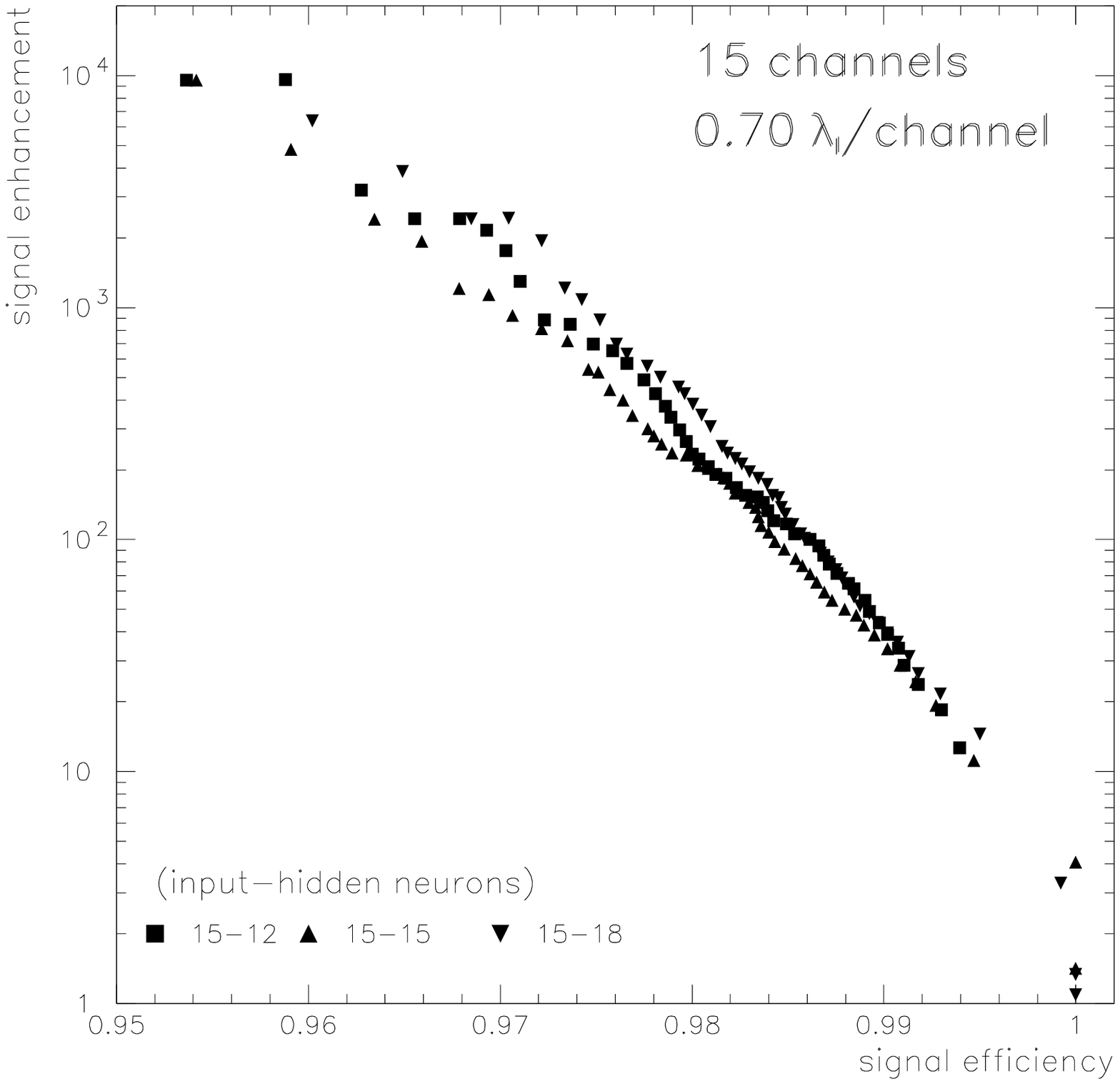}
\caption{signal enhancement ($\frac{\epsilon_{s}}{\epsilon_{b}}$) as a function of signal efficiency
($\epsilon_{s}$) for a calorimeter with 15~channels/octant and total depth of 10.50~$\lambda_{I}$'s.}
\label{fig:sumnnall15}
\end{figure}

\begin{figure}[tbp]
\centering
\epsfxsize=270pt
\epsffile{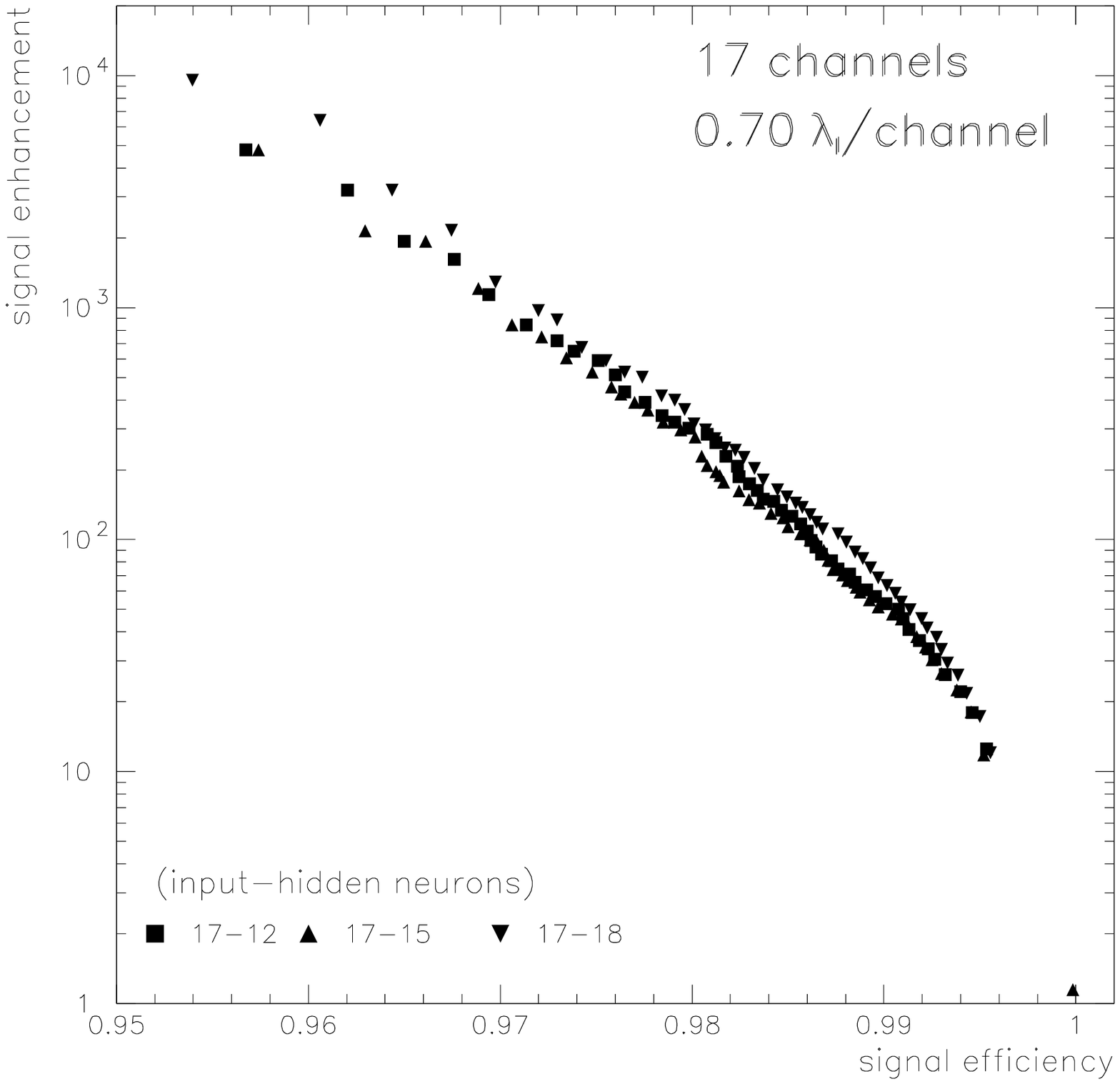}
\caption{signal enhancement ($\frac{\epsilon_{s}}{\epsilon_{b}}$) as a function of signal efficiency
($\epsilon_{s}$) for a calorimeter with 17~channels/octant and total depth of 11.90~$\lambda_{I}$'s.}
\label{fig:sumnnall17}
\end{figure}

In this case each channel consists of 10 consecutive W-fiber layers, which
correspond to depth of 0.70~$\lambda_{I}$ per channel ($= 19.2~X_0$/channel). The calorimeters that have been
studied are composed of 13, 15 and 17 channels per octant (total depth is 9.10, 10.50, 11.90 $\lambda_{I}$'s, respectively).
The same procedure as in the previous case has been followed, neural networks with various numbers of
hidden neurons have been studied.
Their performance is shown in figures~\ref{fig:sumnnall13},~\ref{fig:sumnnall15}~and~\ref{fig:sumnnall17}.
We can still achieve high signal enhancement and efficiency.
Although the channel depth of this case is reduced by 28.6\% compared to the previous one,
we observe that the performance is not significantly improved.
The results from each neural network configuration are presented analytically in
table~\ref{tab:NNs_performance} at page~\pageref{tab:NNs_performance} and
discussed in subsection~4.5.


\subsection{Calorimeter of 0.49 $\lambda_{I}$ per channel}

\begin{figure}[tbp]
\centering
\epsfxsize=270pt
\epsffile{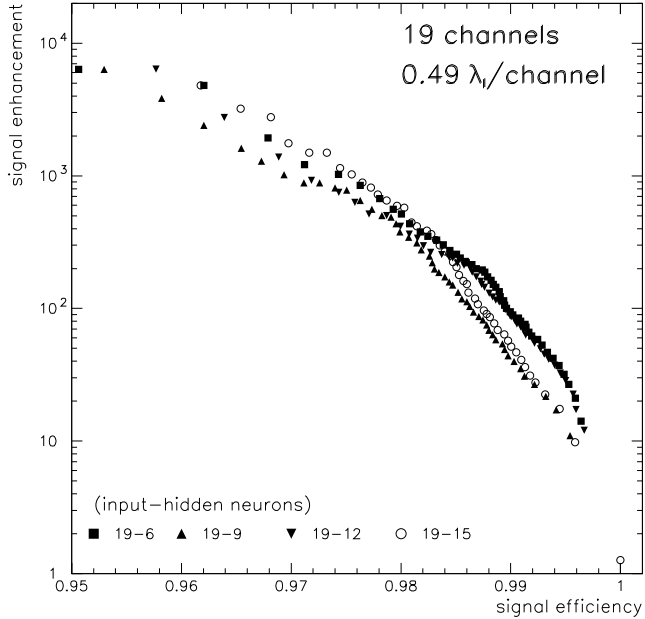}
\caption{signal enhancement ($\frac{\epsilon_{s}}{\epsilon_{b}}$) as a function of signal efficiency
($\epsilon_{s}$) for a calorimeter with 19~channels/octant and total depth of 9.31~$\lambda_{I}$'s.}
\label{fig:sumnnall19}
\end{figure}

\begin{figure}[tbp]
\centering
\epsfxsize=270pt
\epsffile{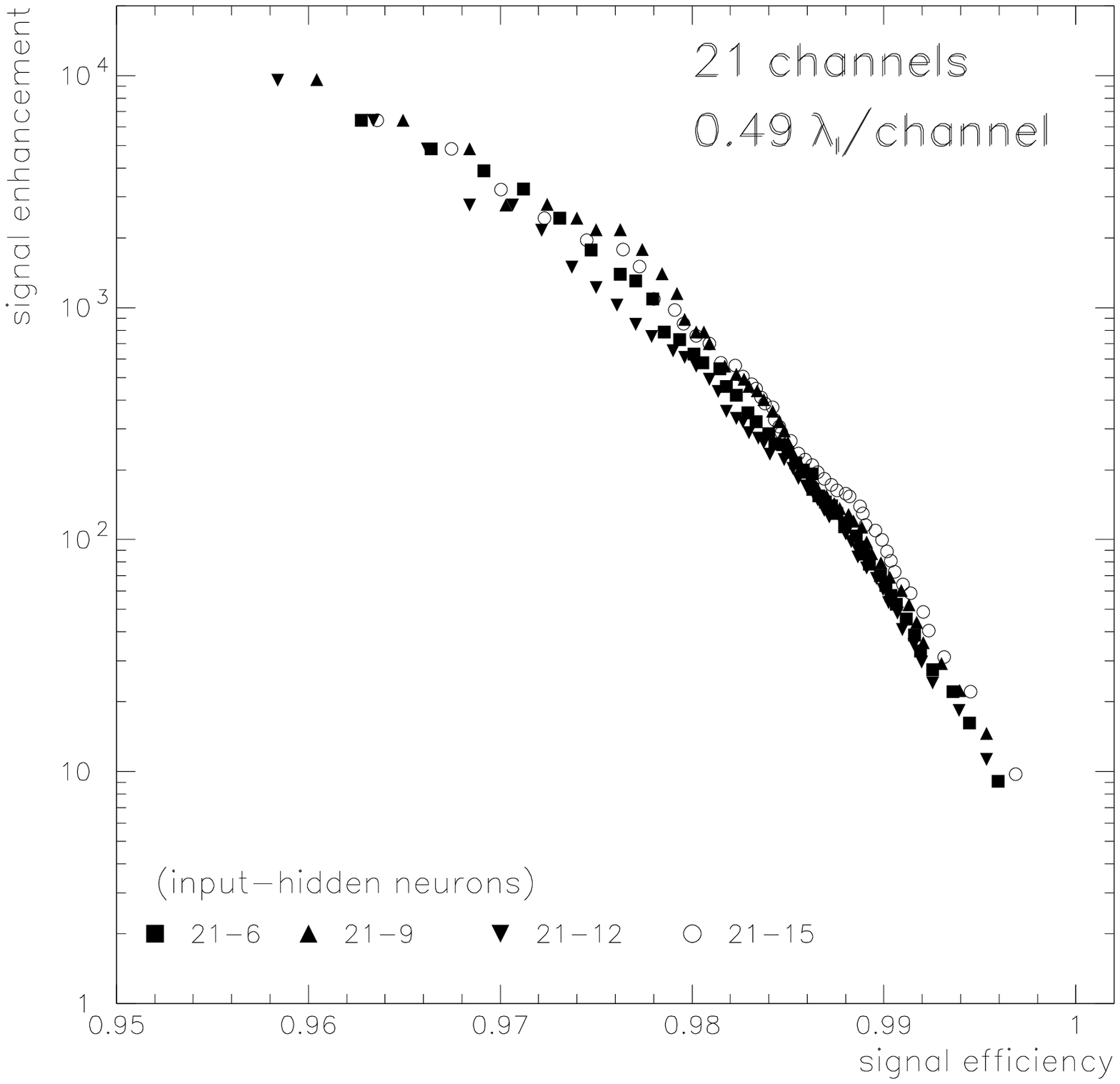}
\caption{signal enhancement ($\frac{\epsilon_{s}}{\epsilon_{b}}$) as a function of signal efficiency
($\epsilon_{s}$) for a calorimeter with 21~channels/octant and total depth of 10.29~$\lambda_{I}$'s.}
\label{fig:sumnnall21}
\end{figure}

\begin{figure}[tbp]
\centering
\epsfxsize=270pt
\epsffile{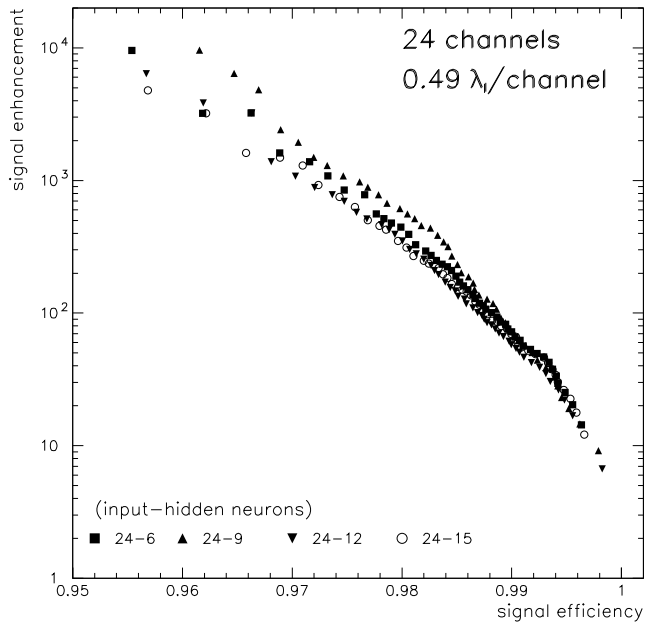}
\caption{signal enhancement ($\frac{\epsilon_{s}}{\epsilon_{b}}$) as a function of signal efficiency
($\epsilon_{s}$) for a calorimeter with 24~channels/octant and total depth of 11.76~$\lambda_{I}$'s.}
\label{fig:sumnnall24}
\end{figure}

We studied also the case where each channel consists of 7 consecutive W-fiber layers,
corresponding to depth of 0.49~$\lambda_{I}$ per channel ($= 13.4~X_0$/channel).
Calorimeters of total depth 9.31, 10.29 and 11.76~$\lambda_{I}$'s have been
studied (composed of 19, 21 and 24 channels per octant, respectively).
A significantly improved performance should be reached to justify the
high cost, due to large number of channels.
We followed the same procedure as in the previous cases.
The performance of the neural networks that have been used is depicted
in figures~\ref{fig:sumnnall19},~\ref{fig:sumnnall21}~and~\ref{fig:sumnnall24}.
The signal enhancement $\frac{\epsilon_{s}}{\epsilon_{b}}$ at fixed efficiency $\epsilon_{s}=0.96$
and the efficiency $\epsilon_{s}$ at fixed $\frac{\epsilon_{s}}{\epsilon_{b}}=1000$ are
presented for each NN architecture in table~\ref{tab:NNs_performance} at page~\pageref{tab:NNs_performance}.
In general, although the performance in terms of achievable signal enhancement and efficiency is very satisfactory,
it is not considerably better compared to that of 1.05 or 0.70~$\lambda_{I}$/channel cases.


In the following we discuss the results of the various calorimeter configurations and neural network architectures
that have been studied.

\subsection{Results recapitulation}

\begin{figure}[f]
\centering
\epsfxsize=300pt
\epsffile{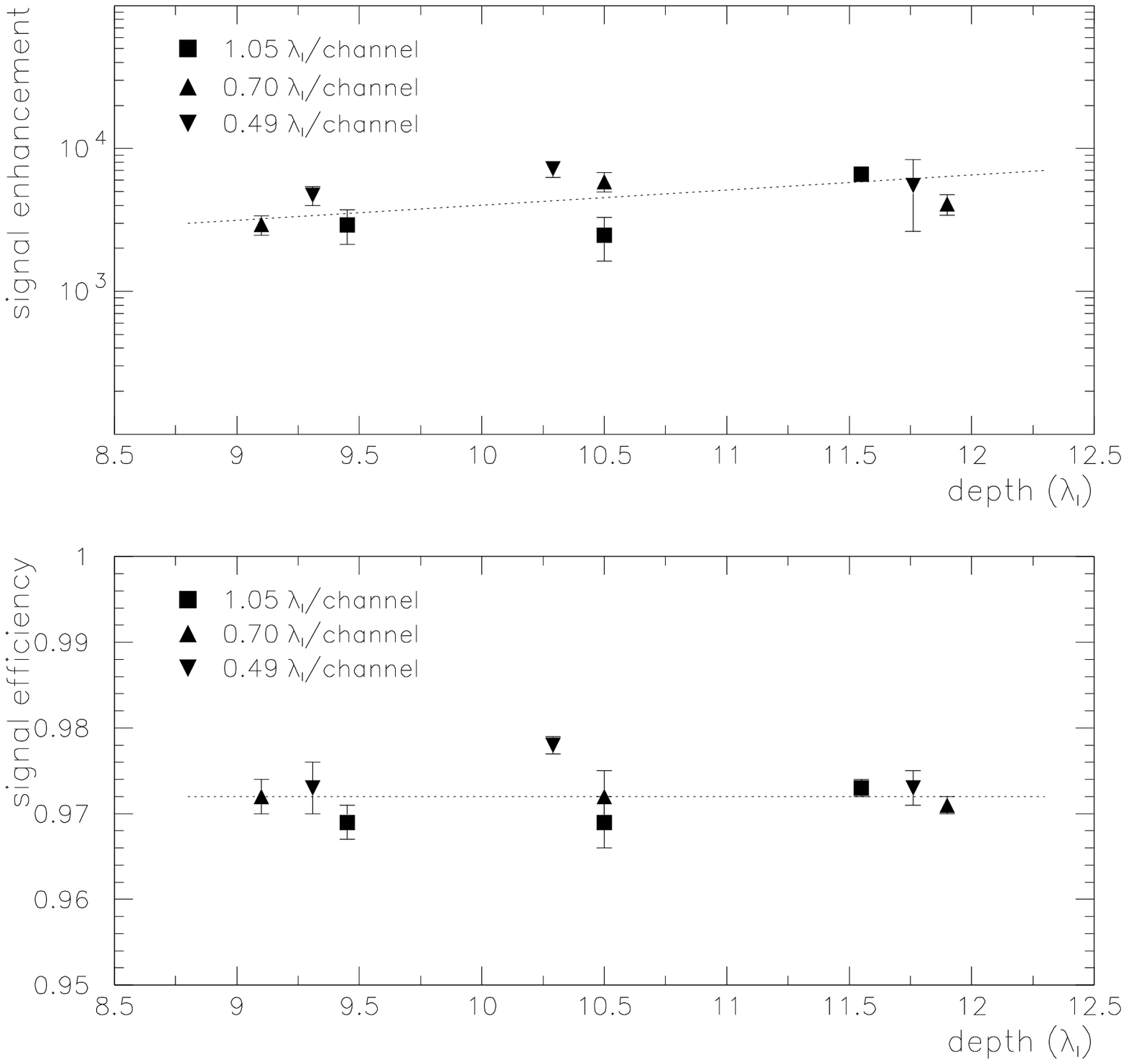}
\caption{average signal enhancement (at $\epsilon_{s}=0.96$) and efficiency (at $\frac{\epsilon_{s}}{\epsilon_{b}}=1000$)
as a function of total calorimeter depth for different
channel configurations. A trendline is shown to guide the eye.}
\label{fig:performance_vs_depth}

\vspace{30pt}

\centering
\epsfxsize=300pt
\epsffile{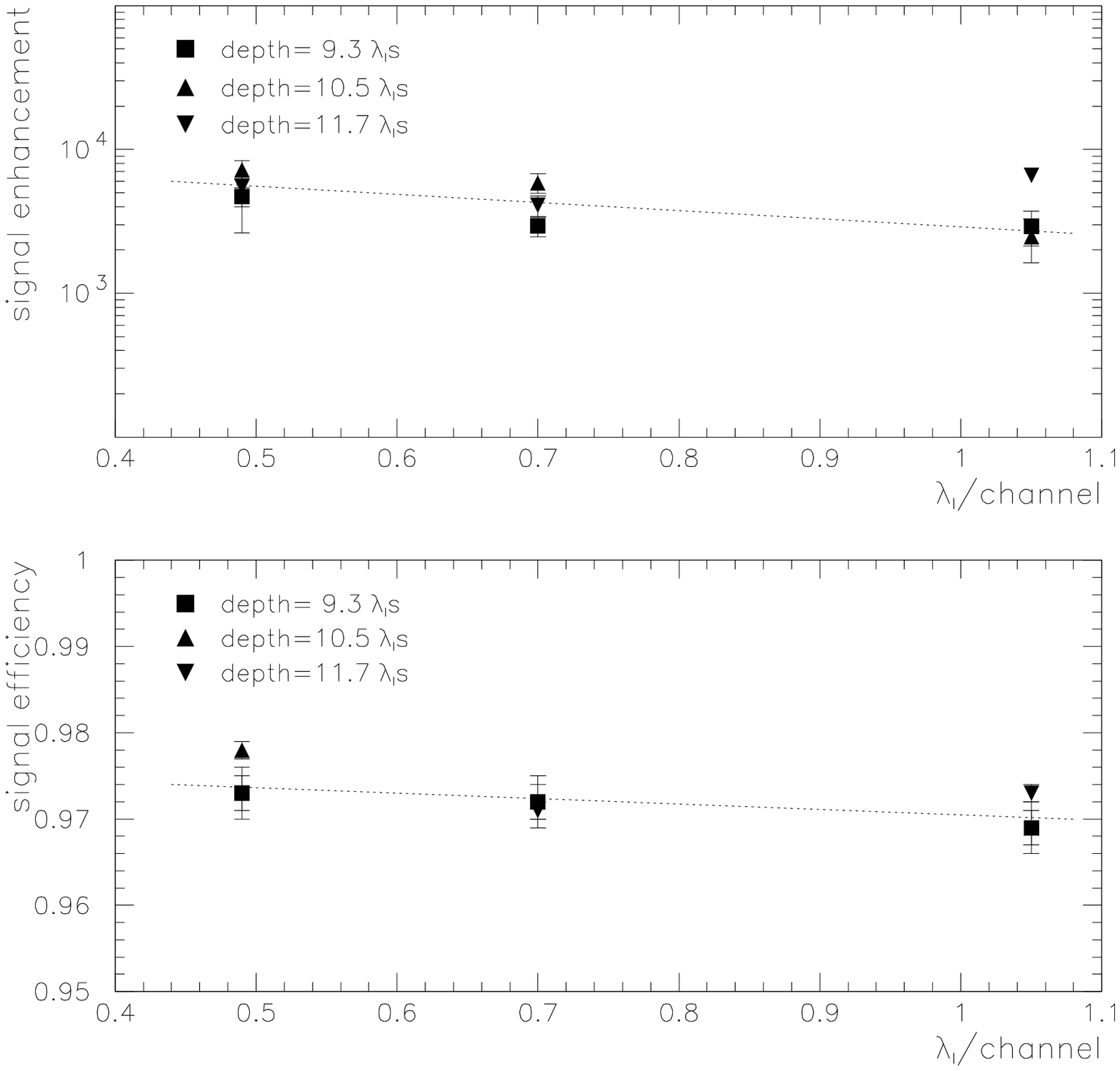}
\caption{average signal enhancement (at $\epsilon_{s}=0.96$) and efficiency (at $\frac{\epsilon_{s}}{\epsilon_{b}}=1000$)
as a function of depth per channel for different
total calorimeter depths. A trendline is shown to guide the eye.}
\label{fig:performance_vs_perchan}
\end{figure}

The signal enhancement $\frac{\epsilon_{s}}{\epsilon_{b}}$ at $\epsilon_{s}=0.96$
and the signal classification efficiency $\epsilon_{s}$ at $\frac{\epsilon_{s}}{\epsilon_{b}}=1000$ are
presented for each neural network architecture and calorimeter configuration in table~\ref{tab:NNs_performance}.
As proposed in~\cite{ref:wolpert92,ref:horn97} we average the performance of the different neural networks which correspond to the same
calorimeter configuration. The results are tabulated in table~\ref{tab:averaged_NNs_performance}
and plotted in figures~\ref{fig:performance_vs_depth}~and~\ref{fig:performance_vs_perchan}
(points are grouped according to depth per channel and total calorimeter depth, respectively).
We observe that the signal enhancement is increasing by increasing calorimeter depth and/or decreasing
depth per channel. The signal efficiency stays basically unchanged, thus the use of a deep calorimeter
with frequent read-out results in more efficient background discrimination (so lower contamination).

Generally, in all calorimeter configurations that have been studied,
the signal-background classification task as performed by neural networks can provide
a signal-over-background enhancement factor, $\frac{\epsilon_{s}}{\epsilon_{b}}$, larger than 2000(1000)
at high signal classification efficiency, $\epsilon_{s}$, of a value of 0.96(0.97).
This performance is considered very satisfactory and can be achieved even with a moderately short calorimeter
(9.45 $\lambda_{I}$'s deep) with a conservative number of channels (9 channels per octant, 1.05~$\lambda_{I}$/channel).
In the case of very frequent read-out (0.49~$\lambda_{I}$/channel),
the performance in terms of achievable signal enhancement and efficiency is improved. But still
is not considerably better, compared to the cases of 1.05 or 0.70~$\lambda_{I}$/channel read-out,
to justify the higher cost of the large total number of calorimeter channels.
If we had to choose a configuration out of the 9 ones that have been studied, we would suggest
either the case of a calorimeter of 11.55~$\lambda_{I}$'s deep with 11 channels per octant (1.05~$\lambda_{I}$/channel)
or one with depth of 10.50~$\lambda_{I}$'s with 15 channels per octant (0.70~$\lambda_{I}$/channel),
since they both combine adequate number of channels and total depth with high performance and cost efficiency.

\begin{table}[f]
\centering
  \caption{signal enhancement $\frac{\epsilon_{s}}{\epsilon_{b}}$ at $\epsilon_{s}=0.96$
and signal efficiency $\epsilon_{s}$ at $\frac{\epsilon_{s}}{\epsilon_{b}}=1000$ for the studied
configurations and neural network architectures.}
  \label{tab:NNs_performance}
\vspace{10pt}
  \begin{tabular}{crrcc}
    \hline\hline

calorimeter& inputs              & hidden & $\frac{\epsilon_{s}}{\epsilon_{b}}$ at $\epsilon_{s} = 0.96$ & $\epsilon_{s}$ at $\frac{\epsilon_{s}}{\epsilon_{b}} = 1000$ \\
read-out   & ($\equiv$ channels) &neurons & & \\
\hline
& & & & \\
1.05 $\lambda_{I}$/channel
& 9   &  3  & 1921 & 0.966 \\
&     &  6  & 3842 & 0.971 \\
&     &  9  & 2800 & 0.969 \\
&     & 12  & 3119 & 0.970 \\
&     &     &      &       \\
& 10  &  6  & 1281 & 0.966 \\
&     &  9  & 2572 & 0.971 \\
&     & 12  & 3197 & 0.971 \\
&     & 15  & 2799 & 0.967 \\
&     &     &      &       \\
& 11  &  3  & 6435 & 0.975 \\
&     &  6  & 7190 & 0.973 \\
&     &  9  & 6720 & 0.973 \\
&     & 12  & 6410 & 0.973 \\
&     & 15  & 6404 & 0.973 \\
&     &     &      &       \\
\hline
&     &     &      &       \\
0.70 $\lambda_{I}$/channel
& 13  &  9  & 3198 & 0.971 \\
&     & 12  & 2404 & 0.972 \\
&     & 15  & 3199 & 0.974 \\
&     &     &      &       \\
& 15  & 12  & 6399 & 0.972 \\
&     & 15  & 4796 & 0.970 \\
&     & 18  & 6401 & 0.975 \\
&     &     &      &       \\
& 17  & 12  & 3996 & 0.970 \\
&     & 15  & 3464 & 0.970 \\
&     & 18  & 4809 & 0.972 \\
&     &     &      &       \\
\hline
&     &     &      &       \\
0.49 $\lambda_{I}$/channel
& 19  &  6  & 5574 & 0.974 \\
&     &  9  & 3832 & 0.970 \\
&     & 12  & 4570 & 0.972 \\
&     & 15  & 4809 & 0.976 \\
&     &     &      &       \\
& 21  &  6  & 6418 & 0.978 \\
&     &  9  & 8019 & 0.979 \\
&     & 12  & 8004 & 0.976 \\
&     & 15  & 6424 & 0.979 \\
&     &     &      &       \\
& 24  &  6  & 3206 & 0.973 \\
&     &  9  & 9615 & 0.975 \\
&     & 12  & 5113 & 0.970 \\
&     & 15  & 3995 & 0.972 \\
& & & \\
\hline\hline
  \end{tabular}
\end{table}

\begin{table}[f]
\centering
  \caption{average signal enhancement ($\frac{\epsilon_{s}}{\epsilon_{b}}$ at $\epsilon_{s}=0.96$)
and average signal efficiency ($\epsilon_{s}$ at $\frac{\epsilon_{s}}{\epsilon_{b}}=1000$)
for the studied calorimeter configurations.}
  \label{tab:averaged_NNs_performance}
\vspace{10pt}
  \begin{tabular}{ccccc}
    \hline\hline

channels & $\lambda_{I}$/channel
                 & depth ($\lambda_{I}$)
                         & $\frac{\epsilon_{s}}{\epsilon_{b}}$ at $\epsilon_{s} = 0.96$
                                               & $\epsilon_{s}$ at $\frac{\epsilon_{s}}{\epsilon_{b}} = 1000$ \\
\hline
         &       &       &                     &                         \\
       9 &  1.05 &  9.45 &  2921 $\pm$ 796     &     0.969 $\pm$  0.002  \\
      10 &  1.05 & 10.50 &  2462 $\pm$ 829     &     0.969 $\pm$  0.003  \\
      11 &  1.05 & 11.55 &  6632 $\pm$ 339     &     0.973 $\pm$  0.001  \\
         &       &       &                     &                         \\
      13 &  0.70 &  9.10 &  2934 $\pm$ 459     &     0.972 $\pm$  0.002  \\
      15 &  0.70 & 10.50 &  5865 $\pm$ 926     &     0.972 $\pm$  0.003  \\
      17 &  0.70 & 11.90 &  4090 $\pm$ 677     &     0.971 $\pm$  0.001  \\
         &       &       &                     &                         \\
      19 &  0.49 &  9.31 &  4696 $\pm$ 718     &     0.973 $\pm$  0.003  \\
      21 &  0.49 & 10.29 &  7216 $\pm$ 918     &     0.978 $\pm$  0.001  \\
      24 &  0.49 & 11.76 &  5482 $\pm$ 2864    &     0.973 $\pm$  0.002  \\
         &       &       &                     &                         \\
\hline\hline
  \end{tabular}
\end{table}

\section{Summary and conclusions}

We presented a signal-from-background separation study based on neural networks technique. We
used a multilayer perceptron with one hidden layer that was fed with
input variables that were the channel responses of each octant of the CASTOR calorimeter.
The network was trained to distinguish between an octant, which contained the
characteristic pattern of the longitudinal development of the shower of a ``long-penetrating object'' (strangelet),
and an octant, which contained only signals from conventional particle showers.

We concentrated on the case of a strangelet with energy of~5~TeV, an amount
which corresponds to 2.5\% of the total energy per event that is expected to be received by the calorimeter.
We studied calorimeter configurations with various total depths and depths per channel.
The results show that we can very efficiently separate the signal from the background and
compensate the initial signal-to-background ratio which is expected to be of the order of 1/10000.
The neural networks based classification task can provide
a signal-over-background enhancement factor larger than 2000(1000),
at signal classification efficiency as high as 0.96(0.97), and thus resulting
to a selected subset of events to be further analyzed with a significantly improved signal-to-background ratio
of the order of 0.1 or higher.
We stress the fact that this performance is achieved without any preprocessing or
preselection procedure, thus an even higher signal-to-background ratio might be accomplished.

Concerning the optimum calorimeter configuration in terms of total depth and read-out frequency, we conclude
that a total depth between 11.55 and 10.50~$\lambda_{I}$'s and with 1.05-0.70~$\lambda_{I}$/channel
(10 to 15 channels per octant) is sufficient to ensure high classification performance.
The fact that such good performance corresponds to a 5~TeV strangelet
leads in concluding that similar or even better performance can be achieved  for strangelets of higher energy
since their signal will be stronger and more pronounced.

\clearpage

{\large\bf References}
\vspace*{-25pt}\\

\end{document}